\documentclass[pra, twocolumn,showpacs,nofootinbibfloatfix,amsmath,amsfonts,amssymb]{revtex4-1}%
\usepackage{amsmath,amsfonts,amssymb,color}
\usepackage{amsthm}
\usepackage{leftidx}
\usepackage{graphicx}
\usepackage{xcolor}
\usepackage{dcolumn}
\usepackage{bm}
\usepackage{epstopdf}
\usepackage{epsfig}
\usepackage{mathdots} 

\def\RBc{\textcolor{black}}

\def\RBB{\textcolor{black}}
\def\RBC{\textcolor{black}}

\newcommand{\IM}{\mathbb I}

\newcommand{\ZM}{\mathbb Z}

\usepackage{caption}
\usepackage{graphicx}
\usepackage{subcaption}
\usepackage{braket}
\usepackage{amssymb}
\usepackage{mathrsfs}
\usepackage{cleveref}
\creflabelformat{equation}{(#2#1#3)}
\crefrangeformat{equation}{(#3#1#4--#5#2#6)}
\usepackage{amssymb}
\newcommand\Tau{\mathcal{T}}
\newcommand\zo{\sigma_z^{(1)}}
\newcommand\zs{\sigma_z^{(2)}}
\newcommand\xo{\sigma_x^{(1)}}
\newcommand\xs{\sigma_x^{(2)}}
\newcommand\yo{\sigma_y^{(1)}}
\newcommand\ys{\sigma_y^{(2)}}
\usepackage{chngcntr}

\begin{document}
	\title{Generating non-Clifford gate operations through exact mapping between Majorana fermions and $\mathbb{Z}_4$ parafermions}
        \author{Ali Hamed Safwan}
        \affiliation{%
		Department of Physics, King Fahd University of Petroleum and Minerals, 31261 Dhahran, Saudi Arabia
	}
	\author{Raditya Weda Bomantara}
	\email{Raditya.Bomantara@kfupm.edu.sa}
	\affiliation{%
		Department of Physics, Interdisciplinary Research Center for Intelligent Secure Systems, King Fahd University of Petroleum and Minerals, 31261 Dhahran, Saudi Arabia
	}
	\date{\today}
	
	
	\vspace{2cm}
	
\begin{abstract}
Majorana fermions and their generalizations to $\mathbb{Z}_n$ parafermions are considered promising building blocks of fault-tolerant quantum computers for their ability to encode quantum information nonlocally. In such topological quantum computers, highly robust quantum gates are obtained by braiding pairs of these quasi-particles. However, it is well-known that braiding Majorana fermions or parafermions only leads to a Clifford gate, hindering quantum universality. This paper establishes an exact mapping between Majorana fermions to $\mathbb{Z}_4$ parafermions in systems under total parity non-conserving and total parity conserving setting. It is revealed that braiding of Majorana fermions may lead to non-Clifford quantum gates in the 4-dimensional qudit representation spanned by $\mathbb{Z}_4$ parafermions, whilst braiding of $\mathbb{Z}_4$ parafermions may similarly yield non-Clifford quantum gates in the qubit representation spanned by Majorana fermions. This finding suggests that topologically protected universal quantum computing may be possible with Majorana fermions ($\mathbb{Z}_4$ parafermions) by supplementing the usual braiding operations with the braiding of $\mathbb{Z}_4$ parafermions (Majorana fermions) that could be formed out of Majorana fermions ($\mathbb{Z}_4$ parafermions) via the mapping prescribed here. Finally, the paper discusses how braiding of Majorana fermions or $\mathbb{Z}_4$ parafermions could be obtained via a series of parity  measurements.

\end{abstract}

\maketitle

\section{Introduction}

Since the end of the last century, the concept of quantum computing has emerged as a paradigmatic idea which utilizes quantum physics machinery to perform computational tasks \cite{Grover1996,Shor1994,Deutsch1985,Feynman1982}. Specifically, rather than storing and processing data as classical bits that consist of $``0"$s and $``1"$s, quantum computing involves the definition and manipulation of quantum bits (qubits), which are quantum states living in a \RBc{$2^n$-dimensional} Hilbert space with $n$ being some positive integer. It has been envisioned that a functional quantum computer is capable of solving tasks that otherwise would take more than a lifetime in under one second \cite{Madsen2022}. 

Building a full-fledged quantum computer is \RBc{no easy feat}. Due to the lack of natural error-correction mechanisms in quantum mechanical systems, qubits are prone to errors that are typically ubiquitous in nature due to, e.g., interaction with environment and/or hardware imperfection \cite{Kitaev2001, Papic2023, Riste2014}. While quantum error-correction algorithms exist and have been extensively studied, their success require an already very small error in the system. For this reason, much effort has been devoted towards designing a physical system that hosts qubits with minimal errors. Topological systems are potentially one of the most promising platforms for this purpose due their natural robustness that results from nonlocality. Indeed, the idea of using topological systems for quantum computing has emerged as an active area on its own which is dubbed as topological quantum computing \cite{Pachos2017,Sarma2015,Fre2008}.   

In topological quantum computing, quantum information is stored nonlocally by pairs of spatially separated non-Abelian anyons, whilst (unitary) quantum gate operations are realized by moving one anyon around another (a process referred to as braiding) \cite{Pachos2017,Sarma2015,Fre2008}. As errors typically occur locally, the quantum information stored in this manner is naturally resistant against such errors. Moreover, since braiding of anyons is achieved without changing the topology of the system, the resulting quantum gate operations are topologically protected against errors that occur during the braiding process \cite{Bonesteel2005}. 

Majorana fermions are a type of non-Abelian anyons that has gained popularity since the work of Ref.~\cite{Kitaev2001}. They are Hermitian operators that mathematically arise as the real and imaginary parts of some fermionic operator, which are thus often regarded as ``half"-fermions. Majorana fermions are particularly attractive for their potential existence in relatively realistic systems, such as in cold-atom setups \cite{jiang2011} or semiconductor-superconductor heterostructures \cite{lutchyn2010, Oreg2010}. Experimental studies of systems that potentially host Majorana fermions have been extensively made since the last decade \cite{Mourik2012,rokhinson2012,NadjPerge2014,Das2012,Chen2017}. In these experiments, the expected signatures of Majorana fermions have been successfully detected, though such signatures could also be attributed to other mechanisms and are not uniquely associated with Majorana fermions alone \cite{Kells2012,Liu2012,Pan2020,Lai2022}. Nevertheless, it is highly likely that the systems considered in these previous experiments support Majorana fermions even though their unambigous detection currently remains an open problem. 

Theoretically, the quantum computing capabilities of Majorana fermions and their scalability have been comprehensively uncovered \cite{Karzig2017,Vijay2015,Vijay2017,Landau2016,Plugge2016}. In particular, quantum gate operations realizable by braiding Majorana fermions are known to belong to the Clifford gate set, which comprises of gates spanned by the Controlled-NOT (CNOT), Hadamard, and the Phase gate \cite{Fre2008}. Unfortunately, gates belonging to the Clifford gate set can be efficiently simulated by a classical computer and thus do not lead to quantum advantage \cite{Bravyi2016}. Efforts toward establishing universality with Majorana fermions involve supplementing a $T$-gate and/or a magic state to the list of gates achievable by braiding. However, such an extra component is typically obtained through a non-topological means, e.g., dynamically \cite{Hyart2013,Freedman2006,Bravyi2006,Bonderson2010} or geometrically \cite{Bomantara2018,Bomantara2018PRB} rendering any resulting non-Clifford gate less robust as compared with the naturally existing Clifford gates.  

$\mathbb{Z}_n$ parafermions are the direct generalizations of Majorana fermions, a pair of which encodes a nonlocal $n$-dimensional qudit rather than a qubit. In particular, whilst a system of Majorana fermions could be mapped to a system of Pauli matrices, a system of $\mathbb{Z}_n$ parafermions is instead mappable to a system of some $n\times n$ generalizations of the Pauli matrices \cite{Fendley2012,Fendley2016}. The quantum computing capabilities of $\mathbb{Z}_n$ parafermions have been systematically studied in Ref.~\cite{Hutter2015}, with the main conclusion being that, like Majorana fermions, braiding of $\mathbb{Z}_n$ parafermions only gives rise to an $n$-dimensional Clifford gate that is not compatible with quantum universality. Nevertheless, even within the Clifford gate set, braiding $\mathbb{Z}_n$ parafermions could potentially lead to richer accessible gates as compared with braiding Majorana fermions \cite{Hutter2015}. Moreover, certain types of parafermions, e.g., $\mathbb{Z}_3$ parafermions, could potentially be building-blocks of a system that harbors the sought-after Fibonacci anyons \cite{Stoudenmire2015}. Here, Fibonacci anyons are a (hypothetical) type of non-Abelian anyons that support non-Clifford gates by braiding operations \cite{Xu2024,Minev2024}.     

Interest in $\mathbb{Z}_n$ parafermions has resulted in extensive theoretical studies that revolve around detecting them in experiments and utilizing them for quantum computations \cite{Liu2022,Kaskela2021,Fendley2012,Clarke2013,Mong2014,Benhemou2022}. While the proposals for realizing general $\mathbb{Z}_n$ parafermions often involve access to the elusive fractional quantum Hall system, restricting ourselves to the specific type of $\mathbb{Z}_4$ parafermions opens some alternative, more realistic, approaches \cite{Chew2018,Calzona2018,Orth2015,Zhang2014}. Indeed, since a mapping exists between $\mathbb{Z}_4$ parafermions and Majorana fermions, see, e.g., Ref.~\cite{Bomantara2021}, $\mathbb{Z}_4$ parafermions could in principle be realized in platforms that support Majorana fermions with the addition of appropriate interaction terms \cite{Chew2018,Calzona2018,Orth2015,Zhang2014} and/or periodic driving \RBc{\cite{Bomantara2021,Sreejith2016,Zhu2022}.}

In this paper, we thoroughly investigate the mapping between $\mathbb{Z}_4$ parafermions and Majorana fermions at the mathematical level, focusing on the minimal numbers of $\mathbb{Z}_4$ parafermions and Majorana fermions for simplicity, both without any constraint and under the conservation of total parity. Specifically, we explicitly write the Majorana fermionic parity, projector, and braiding operators in terms of $\mathbb{Z}_4$ parafermions. Our main finding is the revelation that, while braiding operators of Majorana fermions and $\mathbb{Z}_4$ parafermions lead only to Clifford gates in their respective representation, i.e., two-qubit and four-dimensional qudit respectively, they may lead to non-Clifford gates in the other representation. That is, a braiding operator involving a pair of Majorana fermions may correspond to a non-Clifford gate operation when it is written in terms of the generalized $4\times 4$ Pauli matrices spanned by the $\mathbb{Z}_4$ parafermions. Similarly, a braiding operator involving a pair of $\mathbb{Z}_4$ parafermions may correspond to a non-Clifford gate operation when it is written in terms of the tensor products between Pauli matrices spanned by the Majorana fermions.   

This paper is organized as follows. In Sec.~\ref{model}, we describe the mathematical definitions that are used in the treatment of Majorana fermions and their generalization to parafermions (focusing on $\mathbb{Z}_4$ parafermions). Specifically, it covers Majorana fermions and parafermions defining operators, parity operators, projectors, measurement operators, braiding operators, as well as their mapping to Pauli matrices and generalized Pauli matrices respectively. Section~\ref{nc}, which is further broken down into three parts, focuses on the mapping between a system of \RBc{four Majorana} fermions and a system two $\mathbb{Z}_4$ parafermions without enforcing a total parity constraint. Section~\ref{MPmapnc} mathematically elucidates the exact mapping between the two systems. Section~\ref{braidMFnc} then exhaustively presents the quantum gate operations that can be achieved by braiding every pair among the \RBc{four Majorana} fermions system from the perspective of Pauli matrices and their $4\times 4$ generalizations. Similarly, Sec.~\ref{braidPFnc} exhaustively presents the quantum gate operations that can be achieved by braiding the pair of $\mathbb{Z}_4$ parafermions from both perspectives. Section~\ref{c}, which is also broken into three parts, develops the exact mapping between a system of \RBc{six Majorana} fermions and a system of four $\mathbb{Z}_4$ parafermions under a conserved total parity. Section~\ref{MPmapnc} mathematically details the exact mapping between the two systems. Section~\ref{braidMFc} then exhaustively presents the quantum gate operations that can be achieved by braiding every pair among the \RBc{six Majorana} fermions while preserving the total parity from the perspective of Pauli matrices and their $4\times 4$ generalizations. Similarly, Sec.~\ref{braidPFc} exhaustively presents the quantum gate operations that can be achieved by braiding every pair among the four $\mathbb{Z}_4$ parafermions from both perspectives. Section~\ref{Mbraiding} elucidates a measurement based protocol to realize the braiding of  Majorana fermions and $\mathbb{Z}_4$ parafermions. Finally, Sec.~\ref{Conc} summarizes the results and outlines future research directions.




\section{Mathematical overview}
\label{model}

\subsection{Overview of Majorana fermions}
Majorana fermion operators satisfy the following defining relations,
\begin{equation}
(\gamma_i)^2=\IM , \quad \gamma_i\gamma_j=-\gamma_j\gamma_i , \quad i \neq j,  \{i,j\} \in \mathbb{N}
\end{equation}
The parity operator of a pair of majorana fermions $\gamma_j$ and $\gamma_k$ is defined as ,
\begin{equation}
P_{jk}^{(MF)}=i\gamma_j\gamma_k ,
\end{equation}
which is unitary, squares to identity, and consequently has two eigenvalues of $\pm 1$ (see Appendix~\ref{app:1}). A projector on the $\pm 1$ eigenstate of the parity can then be written as
\begin{equation}
p_{jk}^{(MF),\pm}=\frac{1}{2}(\IM\pm P_{jk}^{(MF)})
\end{equation}

A measurement of a Majorana parity is an act of projecting the system state onto a definite eigenstate of the Majorana parity. The eigenstate on which the state ends up being after an act of measurement is random with some probability. Mathematically, a Majorana parity measurement can be represented by the operator
\begin{equation}
M_{jk}^{(MF)} = p_{jk}^{(MF),s}, \quad s = \begin{cases} 
1 & \text{with some probability} \\ 
-1 & \text{with some probability} 
\end{cases} \label{mparitym}
\end{equation}

A braiding operator $U_{i,j}^{(MF)}$ between a pair of Majorana fermions $\gamma_i$ and $\gamma_j$ is a unitary operator that exchanges each other while leaving other Majorana operators intact, i.e., 

\begin{eqnarray}
\left(U_{i,j}^{(MF)}\right)^\dagger \gamma_j U_{i,j}^{(MF)} = \gamma_i &,&  \left(U_{i,j}^{(MF)}\right)^\dagger \gamma_i U_{i,j}^{(MF)} = -\gamma_j, \nonumber \\
\quad \left(U_{i,j}^{(MF)}\right)^\dagger \gamma_k U_{i,j}^{(MF)} = \gamma_k && (\forall k \notin \{i,j\}).
\end{eqnarray}
It can be written as \RBc{\cite{Ivanov2001,Kauffman2004,Bravyi2002,Padmanabhan2025}} :
\begin{equation}
U_{i,j}^{(MF)} = \frac{1}{\sqrt{2}} (\IM + \gamma_i \gamma_j) 
\end{equation}




Finally, a set of mutually anticommuting Majorana fermion operators can be mapped onto a set of Pauli matrices via the Jordan-Wigner transformation \cite{Fradkin1980},

\begin{equation}
\gamma_{2j-1} = \sigma_y^{(j)} \prod_{i<j} \sigma_z^{(i)}, \quad \gamma_{2j} = \sigma_x^{(j)} \prod_{i<j} \sigma_z^{(i)} \label{JW}
\end{equation}

\subsection{Overview of $\mathbb{Z}_n$ parafermions} 
\label{pararev}

$\ZM_n$ parafermions are generalizations of Majorana fermions that satisfy the following defining relations,
\begin{equation}
(\psi_i)^n=\IM , \quad \psi_i\psi_j=\omega^{sgn(j-i)}\psi_j\psi_i , \quad i \neq j,  \{i,j\} \in \mathbb{N}
\end{equation}
where $\omega=\exp{\frac{2\pi i}{n}}$. Letting $W_n=\{\omega^p, \forall p \in \mathbb{N}, p \leq n \}$,
the parity operator of a pair of $\ZM_n$ parafermions $\psi_i$ and $\psi_j$ is defined as 
\begin{equation}
P_{jk}^{(P)}=\omega^{-\frac{n-1}{2}}\psi_j^\dagger\psi_k ,\quad j \leq k ,
\end{equation}
which is a unitary operator, equal to identity only when raised to the power of $n$, and consequently has a set of eigenvalues equal to $W_n$ (see Appendix~\ref{app:2}). Other definitions presented below specialize to the case of $\ZM_4$, which is the main subject of this manuscript. The eigenstate associated with the $k\in W_n$ eigenvalue of \RBC{the} parity operator will be denoted by $\ket{k}$ in the following. A projector on $\ket{s}$ can then be written as
\begin{equation}
p_{jk}^{(P),s}=\frac{1}{4}\left[\IM+s^3 P_{jk}^{(P)}+s^2 \left(P_{jk}^{(P)}\right)^2+s \left(P_{jk}^{(P)}\right)^3\right], \label{pparitym}
\end{equation}
where $s \in \{\pm 1, \pm i$\}. Similar to the Majorana fermions' case, a measurement of parafermion parity could also be made to project the system onto a random parity eigenstate with some probability.

Unlike Majorana operators, it is impossible to construct a unitary operator that simply exchanges $\psi_i \leftrightarrow \psi_j$ (see Appendix~\ref{app:A}). Therefore, a braiding operator $U_{ij}$ between a pair of parafermion operators is instead defined to satisfy the braiding relations \cite{Fre2008}, 

\begin{equation}
U_{ij}U_{kl}=U_{kl}U_{ij} \text{$\forall$ distinct i,j,k,l}
\end{equation}
\begin{equation}
U_{ij}U_{jl}U_{ij}=U_{jl}U_{ij}U_{jl} 
\end{equation}
For the case of $\mathbb{Z}_4$ parafermions, such a braiding operator can be written as:
\begin{equation}
U_{jk}^{(P)} = \frac{1}{2} (\IM + \psi_k^3\psi_j+ \psi_k^2\psi_j^2+ \psi_k\psi_j^3) \label{braidpf}
\end{equation}
which acts on the affected parafermions as 

\begin{eqnarray}
   \left(U_{jk}^{(P)}\right)^\dagger \psi_j U_{jk}^{(P)} =-i{\psi_k}^3{\psi_j}^2 & \& & \left(U_{jk}^{(P)}\right)^\dagger \psi_k U_{jk}^{(P)} ={\psi_j} . \nonumber \\
\end{eqnarray}

A set of $\ZM_n$ parafermions can be mapped onto a set of generalized Pauli matrices via the Fradkin-Kadanoff transformations \cite{Fradkin1980}:

\begin{equation}
\psi_{2j-1} = \Tau_y^{(j)} \prod_{i<j} \Tau_z^{(i)}, \quad \psi_{2j} = \Tau_x^{(j)} \prod_{i<j} \Tau_z^{(i)} \label{FK}
\end{equation}
Where $\Tau$ denotes the generalized Pauli Matrices of dimension $n$ and $\Tau_y=\omega^\frac{n-1}{2} \Tau_x \Tau_z$. For the case of $n=4$, $\Tau_x$ and $\Tau_y$ take the explicit form,
\[
\Tau_x = 
\begin{pmatrix}
0 & 1 & 0 & 0 \\
0 & 0 & 1 & 0 \\
0 & 0 & 0 & 1 \\
1 & 0 & 0 & 0
\end{pmatrix}, \quad
\Tau_z = \begin{pmatrix}
1 & 0 & 0 & 0 \\
0 & i & 0 & 0 \\
0 & 0 & -1 & 0 \\
0 & 0 & 0 & -i
\end{pmatrix}
\]
A unitary operator acting on an $N$-dimensional ($N$D) qudit is said to be a Clifford operator if it transforms Pauli gates, under conjugation, into Pauli gates. Mathematically this group of operators can be expressed as \cite{Tolar},
\[ C_N=\{ A \in U(N) | AH(N)A^\dagger \in H(N) \}\]
where $U(N)$ denotes the set of all unitary operators of dimension $N$, and $H(N)$ denotes the Weyl-Heisenberg group of $N$ dimension. For even $N$, $H$ can be expressed as:
\[ H(N)= \{ \omega^{\frac{i}{2}}\Tau_x^j\Tau_Z^k | i,j,k \in \{0,1,2,...,N-1\}  \}\]
For the special case of $2^n$ dimensional qubits, Clifford gates include a multiplicative combination of the $S$ gate, Hadamard gate, and CNOT gate \cite{Gottesman}.
An $N$D generalization of the $S$ and Hadamard gates can be written as, respectively, \cite{Tolar}
\begin{equation}
    S_N= \mathrm{diag}(d_0,d_1,d_2,\dots,d_N),
\end{equation}
and 
\begin{equation}
(H_N)_{ij}=\frac{\omega^{jk}}{\sqrt{N}},     
\end{equation} 
where $d_i=\omega^{\frac{i(n-i)}{2}}$ for even $N$ and $i \in \{0,1,2,...,N-1\}$. In particular, 
\[
S_4 = 
\begin{pmatrix}
1 & 0 & 0 & 0 \\
0 & i^{3/2} & 0 & 0 \\
0 & 0 & -1 & 0 \\
0 & 0 & 0 & i^{3/2}
\end{pmatrix}, \quad
H_4 =\frac{1}{2} \begin{pmatrix}

1 & 1 & 1 & 1 \\
1 & i & -1 & -i \\
1 & -1 & 1 & -1 \\
1 & -i & -1 & i
\end{pmatrix}
\]

\section{Quantum gate operations achieved by braiding without parity conservation constraint}
\label{nc}
For simplicity, we consider the minimal system of either \RBc{four Majorana} fermions or two $\mathbb{Z}_4$ parafermions. These \RBc{four Majorana} fermions are labeled as $\gamma_1$, $\gamma_2$, $\gamma_3$, and $\gamma_4$, whilst the equivalent \RBc{two parafermions} are $\psi_1$ and $\psi_2$. In the absence of parity conservation constraint, the Hilbert space spanned by the Majorana fermions is four dimensional and could be described by the following effective Pauli matrices,
\begin{eqnarray}
    \sigma_z^{(1)}=i\gamma_1\gamma_2 &,& \sigma_x^{(1)}=\gamma_2 ,\nonumber \\
    \sigma_z^{(2)}=i\gamma_3\gamma_4 &,& \sigma_x^{(2)}=i\gamma_1\gamma_2\gamma_4 . \label{Paulinc}
\end{eqnarray}
\RBc{Intuitively, $\sigma_z^{(1)}$ ($\sigma_z^{(2)}$) is related to the occupancy of the number operator associated with a full (complex) fermion formed by $\gamma_1$ and $\gamma_2$ ($\gamma_3$ and $\gamma_4$). In this case, the four dimensional Hilbert space could physically represent a system of two complex fermions, one of which is formed by $\gamma_1$ and $\gamma_2$, whilst the other is formed by $\gamma_3$ and $\gamma_4$.}

Equivalently, the same four dimensional Hilbert space could also be spanned by the $4\times 4$ generalized Pauli matrices $\mathcal{T}_x$ and $\mathcal{T}_y$ as defined in Sec.~\ref{pararev}, which could be expressed in terms of two $\mathbb{Z}_4$ parafermions (labelled $\psi_1$ and $\psi_2$) as
\begin{eqnarray}
    \mathcal{T}_z = i^{-\frac{3}{2}} \psi_1^3 \psi_2 &,& \mathcal{T}_x = \psi_1 . \label{Genpaulinc}
\end{eqnarray}

\subsection{Mapping between Majorana fermions and $\mathbb{Z}_4$ parafermions}
\label{MPmapnc}
We note that the set of Pauli matrices of Eq.~(\ref{Paulinc}) could be related to the set of 4D generalized Pauli matrices of Eq.~(\ref{Genpaulinc}) via
\begin{eqnarray}
    \mathcal{T}_z &=& \sigma_z^{(2)} \left(\frac{1+\sigma_z^{(1)}}{2}+\mathrm{i}\frac{1-\sigma_z^{(1)}}{2}\right) , \nonumber \\
    \mathcal{T}_x &=& \frac{\sigma_x^{(1)}+\mathrm{i}\sigma_y^{(1)}}{2}+\sigma_x^{(2)}\frac{\sigma_x^{(1)}-\mathrm{i}\sigma_y^{(1)}}{2} , \label{gpmtopm}
\end{eqnarray}
and
\begin{eqnarray}
    \sigma_z^{1} = (\Tau_z)^2 &, &   \sigma_x^{1} = \frac{1}{2}[(I+\Tau_z^2)\Tau_x+(I-\Tau_z^2)\Tau_x^3] \nonumber \\
     \sigma_x^{2} = (\Tau_x)^2  &, & \sigma_z^{2} =\frac{1}{2}[(1-i)\Tau_z+(1+i)\Tau_z^3] . \label{pmtogpm}
\end{eqnarray}
one could in principle express all operations involving Majorana fermions in terms of parafermions and vice versa. Indeed, we find 
\begin{eqnarray}
\psi_1&=&\frac{1}{2}({i\gamma_1+\gamma_2-i\gamma_1\gamma_4+\gamma_2\gamma_4}) , \nonumber \\
\psi_2&=&{\frac{1+i}{2\sqrt{2}}[(\gamma_2\gamma_3+\gamma_1\gamma_3\gamma_4)-i(\gamma_1\gamma_3+\gamma_2\gamma_3\gamma_4)]} . \nonumber \\
\end{eqnarray}
The corresponding parafermionic parity operator then reads
\begin{equation}
    P_{12}^{(P)}=\frac{i^{3/2}}{\sqrt{2}}(\gamma_3\gamma_4+\gamma_1\gamma_2\gamma_3\gamma_4)=\frac{-1+i}{2}(\gamma_3\gamma_4+\gamma_1\gamma_2\gamma_3\gamma_4) ,
\end{equation}
and the projector is
\begin{eqnarray}
p_{12}^{(P),s}&=& \frac{1}{4}(\IM+s^2i\gamma_1\gamma_2+\frac{1+i}{2}s(s^2+i)\gamma_3\gamma_4 \nonumber \\
&+& \frac{1+i}{2}s(i-s^2)\gamma_1\gamma_2\gamma_3\gamma_4) ,
\end{eqnarray}
Similarly, we further obtain the Majorana operators in terms of parafermions as

\begin{eqnarray}
    \gamma_1 &=& \frac{i}{2}({(\psi_1^3-\psi_1)+\psi_2^2(\psi_1^3+\psi_1)}) , \nonumber \\
    \gamma_2 &=& {\frac{1}{2}({(\psi_1^3+\psi_1)+\psi_2^2(-\psi_1^3+\psi_1)})} , \nonumber \\
    \gamma_3&=&\frac{1}{\sqrt{2}}({i\psi_2\psi_1+\psi_2^3\psi_1^3}) , \nonumber \\
    \gamma_4&=& {-\psi_2^2} .
\end{eqnarray}
Consequently, the Majorana fermion parity operators can be written in terms of parafermions as 
\begin{eqnarray}
P_{12}^{(MF)} &=&  -\psi_2^2\psi_1^2  , \nonumber \\
P_{13}^{(MF)} &=&  \frac{1+i}{2\sqrt{2}}(-\psi_2^3+i\psi_2+i(-\psi_2^3+i\psi_2)\psi_1^2) , \nonumber \\
P_{14}^{(MF)} &=&  -\frac{1}{\sqrt{2}}(\psi_1^3+\psi_1+(\psi_1^3-\psi_1)\psi_2^2) , \nonumber \\
P_{23}^{(MF)} &=&  \frac{1+i}{2\sqrt{2}}(-i\psi_2^3+\psi_2-(\psi_2^3+i\psi_2)\psi_1^2) , \nonumber \\
P_{24}^{(MF)} &=&  \frac{i}{\sqrt{2}}(\psi_1-\psi_1^3+(\psi_1+\psi_1^3)\psi_2^2) , \nonumber \\
P_{34}^{(MF)} &=&  \frac{1}{\sqrt{2}}(-\psi_2^3\psi_1+i\psi_2\psi_1^3) .
\end{eqnarray}
whilst the projectors are 
\begin{eqnarray}
p_{12}^{(MF),\pm} &=& \frac{1}{2}(\IM\mp \psi_2^2\psi_1^2), \nonumber \\
p_{13}^{(MF),\pm} &=&  \frac{1}{2}(\IM\pm\frac{1+i}{2\sqrt{2}}(-\psi_2^3+i\psi_2+i(-\psi_2^3+i\psi_2)\psi_1^2)), \nonumber \\
p_{14}^{(MF),\pm} &=& \frac{1}{2}(\IM\mp\frac{1}{\sqrt{2}}(\psi_1^3+\psi_1+(\psi_1^3-\psi_1)\psi_2^2)) , \nonumber \\
p_{23}^{(MF),\pm} &=&  \frac{1}{2}(\IM\pm \frac{1+i}{2\sqrt{2}}(-i\psi_2^3+\psi_2-(\psi_2^3+i\psi_2)\psi_1^2)) , \nonumber \\
 p_{24}^{(MF),\pm} &=&  \frac{1}{2}(\IM\pm\frac{i}{\sqrt{2}}(\psi_1-\psi_1^3+(\psi_1+\psi_1^3)\psi_2^2)) , \nonumber \\
p_{34}^{(MF),\pm} &=&  \frac{1}{2}(\IM\pm\frac{1}{\sqrt{2}}(-\psi_2^3\psi_1+i\psi_2\psi_1^3)). 
\end{eqnarray}

\subsection{Braiding Majorana Fermions}
\label{braidMFnc}

There are six possible braidings with \RBc{four Majorana} fermions, they are
\begin{eqnarray}
    U_{1,2}^{(MF)} &=& \frac{\IM +\gamma_1\gamma_2}{\sqrt{2}} = \frac{\IM - \mathrm{i} \sigma_z^{(1)}}{\sqrt{2}}= \frac{1-i}{\sqrt{2}} \cdot (S^{(1)})  , \nonumber \\
    U_{1,3}^{(MF)} &=& \frac{\IM +\gamma_1\gamma_3}{\sqrt{2}} =\frac{\IM + i\sigma_y^{(2)} \sigma_x^{(1)}}{\sqrt{2}}  \nonumber \\
    &=& (S^{(2)})^2 \cdot CNOT_{2,1}\cdot  H^{(2)} \cdot CNOT_{2,1}  ,\nonumber \\
    U_{1,4}^{(MF)} &=& \frac{\IM + \gamma_1 \gamma_4}{\sqrt{2}} = \frac{\IM + \mathrm{i} \sigma_x^{(1)} \sigma_x^{(2)}}{\sqrt{2}} \nonumber \\
    &=& (S^{(2)})\cdot CNOT_{2,1} \cdot H^{(2)} \cdot CNOT_{2,1} \cdot (S^{(2)}) , \nonumber \\
    U_{2,3}^{(MF)} &=& \frac{\IM + \gamma_2 \gamma_3}{\sqrt{2}} = \frac{\IM - \mathrm{i} \sigma_y^{(1)} \sigma_y^{(2)}}{\sqrt{2}} \nonumber \\
    &=& S^{(1)}\cdot CNOT_{2,1} \cdot H^{(2)}\cdot CNOT_{2,1} \cdot (S^{(2)})^{2} \nonumber \\
    &\cdot&  (S^{(1)})^{3} , \nonumber \\
    U_{2,4}^{(MF)} &=& \frac{\IM + \gamma_2 \gamma_4}{\sqrt{2}} = \frac{\IM - \mathrm{i} \sigma_y^{(1)} \sigma_x^{(2)}}{\sqrt{2}} \nonumber \\
    &=& CNOT_{1,2}  \cdot H^{(1)} \cdot CNOT_{1,2} \cdot (S^{(1)})^2 , \nonumber \\
    U_{3,4}^{(MF)} &=& \frac{\IM + \gamma_3 \gamma_4}{\sqrt{2}} = \frac{\IM - \mathrm{i} \sigma_z^{(2)}}{\sqrt{2}}=\frac{(1-i)}{\sqrt{2}} (S^{(2)}) .
\end{eqnarray}
It is worth noting that all braiding operators above result in a quantum gate operation that belongs to the Clifford gate set, as expected. Consequently, braiding operations involving Majorana fermions do not lead to quantum universality \cite{Fre2008}. However, it is interesting to note that, when viewed from a 4D qudit perspective, the braiding operators above may not belong to the (four-dimensional) Clifford gate set. For example, we find that
\begin{equation}
    U_{3,4}^{(MF)} = S_4^\dagger \cdot H_4^2\cdot \Tau_z^3\cdot S_4^6 \cdot R_4(-3\pi/4) \cdot H_4^2 ,
\end{equation}
where $R_4(\theta)=\mathrm{diag}(1,e^{i \theta},e^{2 i\theta},e^{3 i \theta})$ is a general 4D qudit rotation, which is generally outside the Clifford gate set. This observation therefore suggests that universal quantum gate operations on 4D qudits spanned by $\mathbb{Z}_4$ parafermions is possible via braiding their Majorana fermionic constituents. 

\subsection{Braiding parafermions}
\label{braidPFnc}

In the minimal system of two $\mathbb{Z}_4$ parafermions, the only possible braiding operator is
\begin{eqnarray}
U^{(P)} &=& \frac{1}{2}(\IM + \psi_2^3 \psi_1 + \psi_2^2 \psi_1^2 + \psi_2 \psi_1^3)=-\frac{1+i}{\sqrt{2}}S_4 . \nonumber \\
\end{eqnarray}
Unsurprisingly, such a braiding leads only to a diagonal quantum gate operation in the qudit space. Interestingly, when viewed from the two-qubit perspective spanned by \RBc{four Majorana} fermions, the resulting quantum gate operation does not belong to the Clifford gate set for qubits, as it contains the sought-after $T$ gate, i.e., 
\begin{equation}
    U^{(P)} = -\frac{1+i}{\sqrt{2}} \sigma_z^{(2)} \cdot \sigma_z^{(1)} \cdot H^{(1)} \cdot CNOT_{2,1} \cdot H^{(1)} \cdot ({T^{(1)}})^\dagger ,
\end{equation}
\RBc{where $T^{(1)}$ is the $T$ gate acting on the first qubit spanned by $\sigma_x^{(1)}$ and $\sigma_z^{(1)}$}. Therefore, conversely to the observation of Sec.~\ref{braidMFnc}, braiding of $\mathbb{Z}_4$ parafermions may also lead to universal gate operations on qubits spanned by Majorana fermions.


\section{Quantum gate operations achieved by braiding under parity conservation constraint}
\label{c}
 Ideally, quantum systems are operated in a closed setting to prevent unwanted interaction effect with the environment. Under this assumption, nontrivial (quasi)particles are not allowed to enter or exit the system. Mathematically, this leads to the conservation of the system's total parity in the case of Majorana or parafermionic systems. By denoting the \RBc{six Majorana} fermions as $\gamma_1$, $\gamma_2$, $\gamma_3$, $\gamma_4$, $\gamma_5$, and $\gamma_6$ and the \RBc{four parafermions} as $\psi_1$, $\psi_2$, $\psi_3$, and $\psi_4$, the conservation of Majorana fermions' total parity can be written as
\begin{equation}
    P_{\rm tot}^{(MF)} =P_{12}^{(MF)} P_{34}^{(MF)} P_{56}^{(MF)} =-i\gamma_1\gamma_2\gamma_3\gamma_4\gamma_5\gamma_6 ,
\end{equation}
whilst the conservation of $\mathbb{Z}_4$ parafermions' total parity reads
\begin{equation}
    P_{\rm tot}^{(P)} = P_{12}^{(P)}P_{34}^{(P)} = i\psi_1^3\psi_2\psi_3^3\psi_4
\end{equation}
These conservation laws respectively reduce the logical Hilbert space of our system of Majorana fermions or parafermions to $4$ dimensions. Consequently, physical operators involving \RBc{six Majorana} fermions, i.e., parities, projectors, and braiding operators, could be written in terms of 4 $\ZM_4$ parafermions and vice versa in such a total parity conserving setting. 

The Hilbert space spanned by the Majorana fermions could be described by the following effective Pauli matrices,
\begin{eqnarray}
    \sigma_z^{(1)}=i\gamma_1 \gamma_2 &,& \sigma_x^{(1)}=i\gamma_1 \gamma_3 ,\nonumber \\
    \sigma_z^{(2)}=i\gamma_4 \gamma_5 &,& \sigma_x^{(2)}=i\gamma_4 \gamma_6 , \label{pmnc}
\end{eqnarray}
\RBc{which physically represents a system of three complex fermions formed by $\gamma_1$ and $\gamma_2$, $\gamma_4$ and $\gamma_5$, and $\gamma_3$ and $\gamma_6$ respectively, under the conservation of the total number of fermions.} It can be easily checked that these operators commute with $P_{\rm tot}^{(MF)}$ and therefore respect the total parity constraint. The same Hilbert space could be equivalently described by the parafermions via the following effective generalized Pauli matrices,
\begin{eqnarray}
    \mathcal{T}_z= i^{-3/2}\psi_1^{3}\psi_2 &,& \mathcal{T}_x = i^{-3/2}\psi_1^{3}\psi_3 , \label{gpmnc}
\end{eqnarray}
both of which commute with $P_{\rm tot}^{(P)}$.

\subsection{Mapping between Majorana fermions and $\mathbb{Z}_4$ parafermions}
\label{MPmapc}
Using Eqs.~(\ref{gpmtopm}), (\ref{pmnc}), and (\ref{gpmnc}), we obtain the following relations between parafermions and Majorana fermions,
\begin{eqnarray}
    i^{-3/2}\psi_1^3 \psi_2 &=& \frac{1+i}{2} i\gamma_4 \gamma_5 - \frac{1-i}{2} \gamma_1 \gamma_2 \gamma_4 \gamma_5 , \nonumber \\
i^{-3/2}\psi_1^3 \psi_3 &=& \frac{i}{2} \gamma_1 \gamma_3 + \frac{1}{2} \gamma_2 \gamma_3 - \frac{1}{2} \gamma_1 \gamma_3 \gamma_4 \gamma_6 - \frac{i}{2} \gamma_2 \gamma_3 \gamma_4 \gamma_6 . \nonumber \\
\end{eqnarray}
By identifying the left hand side as some parafermion parity operators, we immediately obtain,
\begin{eqnarray}
P_{12}^{(P)}&=&\frac{i-1}{2} (\gamma_4 \gamma_5+\gamma_1 \gamma_2\gamma_4\gamma_5) , \nonumber \\
P_{13}^{(P)}&=&\frac{1}{2}(\gamma_2 \gamma_3-\gamma_1 \gamma_3\gamma_4 \gamma_6+i(\gamma_1 \gamma_3-\gamma_2 \gamma_3\gamma_4 \gamma_6)) .\nonumber \\
\end{eqnarray}
The corresponding projectors then read
\begin{eqnarray}
p_{12}^{(P),s}&=&\frac{1}{4}(\IM+s^2i\gamma_1\gamma_2+\frac{1+i}{2}(s+is^3)\gamma_4\gamma_5) \nonumber \\
&-& \frac{1+i}{2}(s-is^3)\gamma_1\gamma_2\gamma_4\gamma_5 , \nonumber \\
p_{13}^{(P),s}&=& \frac{1}{4}(\IM+s^2i\gamma_4\gamma_6+\frac{i}{2}(s+s^3)\gamma_1\gamma_3+\frac{1}{2}(s-s^3)\gamma_2\gamma_3 \nonumber \\ &-&\frac{1}{2}(s+s^3)\gamma_1 \gamma_3\gamma_4 \gamma_6-\frac{i}{2}(s-s^3)\gamma_2 \gamma_3\gamma_4 \gamma_6) 
\end{eqnarray}
Using Eqs.~(\ref{pmtogpm}), (\ref{pmnc}), and (\ref{gpmnc}), we further obtain the following relations
\begin{eqnarray}
    i\gamma_1 \gamma_2&=&(i^{-3/2}\psi_1^3\psi_2)^2 , \nonumber \\ 
    i\gamma_1 \gamma_3&=& \frac{1}{2}[(I+(i^{-3/2}\psi_1^3\psi_2)^2)(i^{-3/2}\psi_1^3\psi_3) \nonumber \\ 
    &+ &(I-(i^{-3/2}\psi_1^3\psi_2)^2)(i^{-3/2}\psi_1^3\psi_3)^3] , \nonumber \\
    i\gamma_4 \gamma_5 &=&\frac{1}{2}[(1-i)(i^{-3/2}\psi_1^3\psi_2)+(1+i)(i^{-3/2}\psi_1^3\psi_2)^3] , \nonumber \\
     i\gamma_4 \gamma_6 &=& (i^{-3/2}\psi_1^3\psi_3)^2 .
\end{eqnarray}
By identifying the left hand side as some Majorana fermion parity operators and using the algebraic properties of parafermion operators, we now obtain their parafermionic representations,
\begin{eqnarray}
    P_{12}^{(MF)}&=& -\psi_1^2\psi_2^2 , \nonumber \\
    P_{13}^{(MF)} &=& -\frac{1+i}{2\sqrt{2}}(\psi_1\psi_3^3+\psi_1^3\psi_3+\psi_1\psi_2^2\psi_3-\psi_1^3\psi_2^2\psi_3^3) , \nonumber \\
    P_{45}^{(MF)} &=& \frac{-1}{\sqrt{2}}(\psi_1^3\psi_2+i\psi_1\psi_2^3) , \nonumber \\
    P_{46}^{(MF)} &=& -\psi_1^2\psi_3^2 .
\end{eqnarray}
The corresponding projectors are 
\begin{eqnarray}
p_{12}^{(MF),s}&=&\frac{1}{2}(\IM-s\psi_1^2\psi_2^2)\nonumber \\ 
p_{13}^{(MF),s} &=& \frac{1}{2}[\IM-s\frac{1+i}{2\sqrt{2}}(\psi_1\psi_3^3+\psi_1^3\psi_3+\psi_1\psi_2^2\psi_3 \nonumber \\
&-&\psi_1^3\psi_2^2\psi_3^3)] , \nonumber \\
p_{45}^{(MF),s} &=& \frac{1}{2}(\IM-s\frac{1}{\sqrt{2}}(\psi_1^3\psi_2+i\psi_1\psi_2^3)) , \nonumber \\
p_{46}^{(MF),s} &=& \frac{1}{2}(\IM-s\psi_1^2\psi_3^2) .
\end{eqnarray}


\subsection{Braiding Majorana fermions}
\label{braidMFc}

There are totally $15$ possible braidings with \RBc{six Majorana} fermions. The six braiding operators from Sec.~\ref{braidMFnc} now correspond to quantum gate operations
\begin{eqnarray}
    U_{1,2}^{(MF)} &=& \frac{\IM +\gamma_1 \gamma_2}{\sqrt{2}} =   \frac{\IM - \mathrm{i} \sigma_z^{(1)}}{\sqrt{2}}= \frac{1-i}{\sqrt{2}} \cdot (S^{(1)}) , \nonumber \\
    U_{1,3}^{(MF)} &=& \frac{\IM +\gamma_1\gamma_3}{\sqrt{2}} =\frac{\IM - \mathrm{i} \sigma_x^{(1)}}{\sqrt{2}} = \frac{(1-i)}{\sqrt{2}}H^{(1)}\cdot S^{(1)}\cdot H^{(1)}  , \nonumber \\
    U_{1,4}^{(MF)} &=& \frac{\IM + \gamma_1 \gamma_4}{\sqrt{2}} = \frac{\IM + \mathrm{i} \sigma_y^{(1)} \sigma_y^{(2)}}{\sqrt{2}} \nonumber \\
    &=& S^{(1)}\cdot CNOT_{2,1} \cdot H^{(2)}\cdot CNOT_{2,1} \nonumber \\
    &\cdot & (S^{(2)})^{2} \cdot (S^{(1)})^{3} ,\nonumber \\
    U_{2,3}^{(MF)} &=& \frac{\IM + \gamma_2 \gamma_3}{\sqrt{2}} =\frac{\IM +  \sigma_z^{(1)}\sigma_x^{(1)}}{\sqrt{2}}= \sigma_z^{(1)}\cdot H^{(1)}  ,\nonumber \\
    U_{2,4}^{(MF)} &=& \frac{\IM + \gamma_2 \gamma_4}{\sqrt{2}} = \frac{\IM -  \sigma_x^{(1)}\sigma_z^{(2)}\sigma_x^{(2)}}{\sqrt{2}}   \nonumber \\ &=& CNOT_{2,1} \cdot H^{(2)} \cdot CNOT_{2,1} \cdot \sigma_z^{(2)}  ,\nonumber \\
    U_{3,4}^{(MF)} &=& \frac{\IM + \gamma_3 \gamma_4}{\sqrt{2}} = \frac{\IM +  \sigma_z^{(1)}\sigma_z^{(2)}\sigma_x^{(2)}}{\sqrt{2}} 
    \nonumber \\ 
    &=&  H^{(1)} \cdot \sigma_z^{(1)} \cdot CNOT_{2,1} \cdot H^{(2)} \cdot CNOT_{2,1} \cdot \sigma_z^{(2)}  \nonumber  \\
    &\cdot&  \sigma_z^{(1)} \cdot H^{(1)} .
\end{eqnarray}

The difference in the resulting quantum gate operation as compared in Sec.~\ref{braidMFnc} when braiding the same pair of Majorana fermions is attributed to the fact that the effective Pauli matrices here take different forms from those of Sec.~\ref{braidMFnc} so that they are compatible with the total parity conservation constraint. Nevertheless, it is worth noting that the quantum gate operations arising from these braiding operations also belong to the Clifford set. For completeness, we present the braiding operators for the remaining eight pairs of Majorana fermions and their corresponding quantum gate operations in Appendix~\ref{app:Majbraid}. 

Even though braiding operations only result in Clifford gates, when viewed from the perspective of 4D qudits that are spanned by the $\mathbb{Z}_4$ parafermions, non-Clifford gate operations may arise. For example, the braiding operator between $\gamma_1$ and $\gamma_3$ maps a $4D$ Pauli gate to a non-Pauli gate according to 
\begin{equation}
 U_{13}^{(MF)} \cdot \Tau_x \cdot \left( U_{13}^\dagger\right)^{(MF)} =\frac{1}{2} (\Tau_x+\Tau_x^3+i\Tau_z^2(\Tau_x-\Tau_x^3)) .   
\end{equation}
\RBc{Therefore, $U_{13}^{(MF)}$ is a non-Clifford qudit gate.}

\subsection{Braiding Parafermions}
\label{braidPFc}
All possible braiding operators achievable with four $\mathbb{Z}_4$ parafermions reads (See Appendix~\ref{app:Pfbraid} for the detailed procedure in obtaining them)
\begin{eqnarray*}
    U_{1,2}^{(P)} &=& -\frac{1+i}{\sqrt{2}}S_4  , \nonumber \\
    U_{1,3}^{(P)} &=& \frac{-1-i}{\sqrt{2}} H_4^3 \cdot \Tau_z \cdot H_4 \cdot \Tau_x \cdot H_4 \cdot S_4 \cdot H_4^3  , \nonumber \\
    U_{1,4}^{(P)} &=& H_4 \cdot \Tau_z^3 \cdot H_4 \cdot S_4^2 \cdot \Tau_z \cdot  H_4  , \nonumber \\
    U_{2,3}^{(P)} &=&   S_4^4 \cdot H_4 \cdot \Tau_z^3 \cdot H_4 \cdot S_4^2 \cdot \Tau_z \cdot H_4 \cdot S_4^4 \\ \nonumber
    U_{2,4}^{(P)} &=& -i \cdot H_4^3 \cdot \Tau_x^3 \cdot (S_4)^{\dagger} \cdot \Tau_x^3 \cdot \Tau_z^3 \cdot H_4^3 , \nonumber \\
    U_{3,4}^{(P)} &=& U_{1,2}^{(P)} .
\end{eqnarray*}
With the availability of more parafermions, entangling qudit gate operations could be obtained. However, it is worth noting that all the achievable gates above belong to the Clifford gate set for a 4D qudit and thereby cannot establish quantum universality. 

Remarkably, when viewed from the two-qubit perspective spanned by 6 Majorana fermions under the total parity conservation constraint, some non-Clifford quantum gate operations can be achieved. Through the mapping 
\begin{eqnarray}
    S_4 &=& \sigma_z^{(2)} \cdot S^{(1)} \cdot T^{(1)} \cdot H^{(1)} \cdot CNOT_{2,1}  \cdot H^{(1)} , \nonumber \\ 
    H_4 &=& CNOT_{2,1} \cdot \sigma_z^{(2)} \cdot (S^{(1)})^3 \cdot \sigma_x^{(2)} \cdot \sigma_x^{(1)} \cdot CNOT_{1,2} \cdot \sigma_x^{(1)} \nonumber \\
    &\cdot& \sigma_z^{(2)} \cdot (S^{(1)})^3 \cdot CNOT_{2,1} 
    \cdot CR(-\frac{3\pi}{4}) \cdot (CS)^{\dagger} \cdot CNOT_{2,1} \nonumber \\
    &\cdot&  CR(-\frac{3\pi}{4}) \cdot  H^{(1)} \cdot H^{(2)} , \nonumber \\
\Tau_x&=&\sigma_x^{(1)}\cdot H^{(2)}\cdot \sigma_z^{(2)}\cdot H^{(2)}\cdot CNOT_{1,2}  \nonumber ,\\
\Tau_z&=&\sigma_z^{(2)}\cdot (S^{(1)}) , \label{PGMP}
\end{eqnarray}
\RBc{non-Clifford gates can be clearly seen in $U_{1,2}$ that contains a $T$ gate ($T^{(1)}$)}, as well as in
\begin{eqnarray}
    U_{2,3}^{(P)} &=& \sigma_x^{(1)} \cdot  CNOT_{2,1} \cdot \sigma_z^{(2)} \cdot (S^{(1)})^3 \cdot \sigma_x^{(2)} \cdot \sigma_x^{(1)}      \nonumber \\  
    &\cdot& CNOT_{1,2} \cdot \sigma_x^{(1)} \cdot \sigma_z^{(2)} \cdot (S^{(1)})^3 \cdot CNOT_{2,1}  \cdot \sigma_x^{(1)}  \nonumber \\  
    &\cdot& \sigma_z^{(1)} \cdot CR\left(\frac{3\pi}{4}\right) \cdot  S^{(2)} \cdot CS \cdot CNOT_{2,1} \nonumber \\
    &\cdot& CR\left(\frac{3\pi}{4}\right)   \cdot H^{(1)} \cdot H^{(2)} . 
\end{eqnarray}
that contains the controlled $S$ gate ($CS$) and a controlled rotation gate by a nontrivial angle via
\begin{equation}
    CR(\theta) = \begin{pmatrix}
1 & 0 & 0 & 0 \\
0 & 1 & 0 & 0 \\
0 & 0 & \cos(\theta) & -\sin(\theta) \\
0 & 0 & \sin(\theta) & \cos(\theta) 
\end{pmatrix}.
\end{equation}
Therefore, even under the conservation of the total parity, universal quantum computing involving Majorana qubits can be achieved by appropriate braiding of parafermions that could be formed out of several Majorana fermions.

\section{Measurement-based braiding protocol}
\label{Mbraiding}
\begin{figure}[h]
    \centering
    \includegraphics[width=0.45\textwidth]{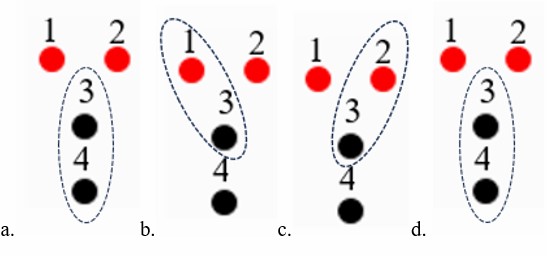}
    \caption{A four-step-protocol to realize a braiding operation by a series of measurements.}
    \label{fig:braidp}
\end{figure}

A promising means to realize the various braiding operations involving pairs of Majorana fermions or parafermions is through devising a series of measurements with the help of some ancilla Majorana fermions or parafermions respectively \cite{Bonderson2013,Bonderson2008,Zheng2016,2Hutter2015}. In both Majorana fermion and parafermion cases, a series of measurements that realize a single braiding operation is schematically depicted in Fig.~\ref{fig:braidp}. There, the red circles denote the main quasiparticles (Majorana fermions or parafermions) to be braided, whereas the black circles denote the ancilla quasiparticles that help facilitate the braiding process. For each step, the parity associated with the quasiparticles inside the dotted ellipse is measured. 

Experimentally, such a parity measurement could be made in at least two different ways. First, a non-superconducting wire could be added to the system that hosts a pair of Majorana fermions whose parity is to be measured. It then follows that, when two terminals are connected to the system, the resulting conductance becomes parity dependent due to the interference effect between the superconducting and the non-superconducting paths \cite{Plugge2017,Bomantara2020,Akhmerov2009}. In this case, measuring the system conductance then yields the effect of measuring the associated Majorana parity as well. Second, one could alternatively couple each of the system's edges that contain Majorana fermions to a common two-level quantum dot, then measure the now parity-dependent state of the quantum-dot-Majorana-fermions composite system \cite{Lutchyn2018,Karzig2017,Bomantara2020s}. Such a measurement could be made via energy level spectroscopy method that amounts to connecting the coupled system to a superconducting transmission line resonator and performing the reflectometry technique to measure the parity-dependent resonator frequency \cite{Karzig2017}. While the above discussion focuses on the parity measurement involving Majorana fermions, it is expected that similar techniques could be adapted and applied to measure the parity of $\mathbb{Z}_4$ parafermions. 

Under a specific set of measurement outcomes, the exact braiding operator is enacted on the subspace spanned by the quasiparticles. However, should the measurement outcomes differ from those required to obtain the target braiding operator, one may either repeat the protocol indefinitely until the intended measurement outcomes are obtained or simply apply appropriate correctional Pauli gates if they are available \cite{Litinski2017}. The mathematical detail of the protocol will be further elaborated in the following for each quasiparticle type. 

\subsection{Measurement-based braiding of Majorana fermions}

Let $\gamma_1$ and $\gamma_2$ be the Majorana fermions that are subject to braiding, while $\gamma_3$ and $\gamma_4$ are the ancillary Majorana fermions. A possible series of measurements that realizes a braiding of $\gamma_1$ and $\gamma_2$ as illustrated in Fig.~\ref{fig:braidp} reads
\begin{equation}
    M^{(MF)}=M^{(MF)}_{34}M^{(MF)}_{23}M^{(MF)}_{13}M^{(MF)}_{34},
\end{equation}
where $M_{ij}^{(MF)}$ is the Majorana parity measurement operator as defined in Eq.~(\ref{mparitym}). By noting that $(\IM + s \gamma_i \gamma_j)(\IM - s' \gamma_i \gamma_j)=2 \delta_{s,s'}(\IM + s \gamma_i \gamma_j)$, we may write
\begin{equation*}
\begin{aligned}
M^{(MF)} &= \frac{1}{16} (\IM + s_1 \gamma_3 \gamma_4)(\IM + s_2 \gamma_2 \gamma_3)(\IM + s_3 \gamma_1 \gamma_3) \\
& \times (\IM + s_4 \gamma_3 \gamma_4) \\
  &= \frac{1}{8} (\IM + s_2 s_3 \gamma_1 \gamma_2)(\IM + s_1 \gamma_3 \gamma_4)\delta_{s_1,s_4} \\
  &+ \frac{1}{8} (s_3 \gamma_1 \gamma_3 + s_2 \gamma_2 \gamma_3)(\IM + s_4 \gamma_3 \gamma_4)\delta_{-s_1,s_4}
\end{aligned}
\end{equation*}
By focusing on the effect on the main subspace spanned by $\gamma_1$ and $\gamma_2$, i.e., ignoring the $\IM \pm \gamma_3 \gamma_4$ factor, this protocol results in
\[U_1^{(MF)}\propto I+\gamma_1\gamma_2 \text{\quad if $s_2=s_3$ and $s_1=s_4$}\]
\[U_2^{(MF)}\propto I-\gamma_1\gamma_2\propto\gamma_2\gamma_1U_1^{(MF)} \text{\quad if $s_2=-s_3$ and $s_1=s_4$}\]
\[U_3^{(MF)}\propto \gamma_1+\gamma_2\propto\gamma_1U_1^{(MF)} \text{\quad   if $ s_2=s_3$ and $s_1=-s_4$}\]
\[U_4^{(MF)}\propto \gamma_1-\gamma_2\propto-\gamma_2U_1^{(MF)} \text{\quad  if $s_2=-s_3$ and $s_1=-s_4$}\]
Regardless of the measurement outcomes, the result is proportional to the true braiding operator $U_1^{(MF)}$ up to products of some Majorana fermions that can be written in terms of appropriate Pauli gates. Therefore, if the Pauli gates are readily accessible in the platform used to enact the series of measurements above, one may simply apply the appropriate Pauli gates depending on the recorded measurement outcomes to retrieve $U_1^{(MF)}$ \cite{P}. 

Alternatively, due to the symmetry in the measurements involved in the protocol, one may in principle repeat the protocol indefinitely and keep track of all the measurement outcomes until exactly $U_1^{(MF)}$ is achieved. \RBc{Assuming all outcomes of the protocol are equally probable, there is a $1/4$ chance that braiding operation is obtained directly in one cycle of measurements. In general, applying the protocol $23$ times ($92$ measurements in total) yields 99.9\% probability (this can be calculated using Binomial Distribution) to achieve the desired braiding. In this case, $92$ is the safe upper bound for the number of measurements needed to realize a braiding operation. In experimental platforms based on InAs/Al heterostructures, recent work \cite{Aghaee2024} reports that a single parity measurement could be completed on a scale of a few microseconds. Therefore, the $92$ measurements could be completed in less than a millisecond, the latter of which is a typical quasiparticle poisoning time of the system \cite{Aghaee2024}.}  

\subsection{Measurement-based braiding of parafermions}
\label{MBPF}
The braiding operator of two $\mathbb{Z}_4$ parafermions could be similarly realized by a series of parafermion measurements with the help of two additional ancilla parafermions. Specifically, let $\psi_1$ and $\psi_2$ be the main parafermions to be braided and $\psi_3$ and $\psi_4$ be the ancillary parafermions. A possible series of measurements that realizes a braiding of $\psi_1$ and $\psi_2$ as illustrated in Fig.~\ref{fig:braidp} now reads
\begin{equation}
    M^{(P)}=M^{(P)}_{34}M^{(P)}_{23}M^{(P)}_{13}M^{(P)}_{34},
\end{equation}
where $M_{ij}^{(P)}$ is the parafermion parity measurement operator, which is given by Eq.~(\ref{pparitym}) but with the measurement outcome $s$ being randomly selected among $1,i,-1,-i$. Through a similar mechanism as in the Majorana fermions case, this protocol results in either a braiding operator or a braiding operator multiplied by a combination of 4D generalized Pauli matrices depending on the measurement outcomes. Specifically, by leaving the mathematical details in Appendix~\ref{app:pmeasure}, we obtain the following effective unitary operator acting on the main subspace for $s_1=s_4$, 
\begin{eqnarray}
U_1^{(P)} &=& \frac{1}{2}(\IM + \psi_2\psi_1^3 + \psi_2^2\psi_1^2 + \psi_2^3\psi_1)   \quad \text{if \(s_3 \cdot s_2^3 = 1 \)} , \nonumber \\
U_i^{(P)} &=&- i\psi_2^3\psi_1 U_1^{(P)} \quad \text{if \(s_3 \cdot s_2^3 = i \)} , \nonumber \\
U_{-1}^{(P)} &=& \psi_2^2\psi_1^2 U_1^{(P)} \quad \text{if \(s_3 \cdot s_2^3 = -1 \)} , \nonumber \\
U_{-i}^{(P)} &=&-  i\psi_2\psi_1^3 U_1^{(P)} \quad \text{if \(s_3 \cdot s_2^3 = -i \)} ,
\end{eqnarray}
and for the remaining values of $s_1$ and $s_4$,
\begin{eqnarray}
   U_2^{(P)} &=& s_2^3 \cdot P_{23}^{(P)} \cdot U_s^{(P)} \quad \text{if \(s_3 \cdot s_2^3 = s \) and \(s_1 = -i s_4 \)} , \nonumber \\
U_3^{(P)} &=& s_2^2 \left(P_{23}^{(P)}\right)^2 \cdot U_s^{(P)} \quad \text{if \(s_3 \cdot s_2^3 = s \) and \(s_1 = - s_4 \)} , \nonumber \\
U_4^{(P)} &=& s_2 \left(P_{23}^{(P)}\right)^3 \cdot U_s^{(P)} \quad \text{if \(s_3 \cdot s_2^3 = s \) and \(s_1 = i s_4 \)} . \nonumber \\
\end{eqnarray}
In particular, we observe that $U_1^{(P)}$ (corresponding to the measurement outcomes of $s_1=s_4$ and $s_3=s_2$) is already the desired braiding operator. The other unitary operators that arise from obtaining the various different measurement outcomes are related to $U_1^{(P)}$ by a product of parafermion operators, which can in turn be expressed in terms of 4D generalized Pauli matrices via the Fradkin-Kadanoff transformation. \RBc{Assuming all outcomes of the protocol are equally probable, applying one cycle of measurements has a $1/16$ chance to yield the desired braiding protocol directly. Applying the protocol $62$ times ($248$ measurements in total) then yields braiding with 98\% probability. Assuming that a single measurement of parafermion parity takes around the same time as a single measurement of Majorana fermion parity, $248$ measurements correspond to a timescale that is of the order of hundreds of microseconds in a typical InAs/Al heterostructures \cite{Aghaee2024}, which is shorter than the quasiparticle poisoning time of the system, i.e., typically of the order of milliseconds \cite{Aghaee2024}.}

\section{Discussion and concluding remarks}
\label{Conc}

In this paper, the exact mapping between Majorana fermions and $\mathbb{Z}_4$ parafermions was developed by focusing on two minimal systems with $4$D Hilbert space. One of these systems consists of either \RBc{four Majorana} fermions or two $\mathbb{Z}_4$ parafermions without any constraint on the total parity, whereas the other consists of either \RBc{six Majorana} fermions or four $\mathbb{Z}_4$ parafermions under a conserved total parity. The main finding of this paper is the revelation that brading of Majorana fermions may correspond to a non-Clifford quantum gate when viewed from the parafermionic 4D qudit perspective, and vice versa. \RBc{In particular, we have shown explicitly that some braiding operators involving Majorana fermions could be decomposed into 4D qudit gates that include non-Clifford components, such as a single qudit rotation by $\theta=-3\pi/4$. Conversely, we have also explicitly demonstrated the decomposition of some braiding operators involving parafermions in terms of non-Clifford qubit gate components, such as a $T$ gate, a controlled $S$ gate, and/or a controlled rotation gate by $\theta=\pm 3\pi/4$.} This result suggests that quantum universality could, in principle, be established in a topological manner in systems of either Majorana fermions or $\mathbb{Z}_4$ parafermions if the braiding operations involving one type of quasiparticles (Majorana fermions or $\mathbb{Z}_4$ parafermions) are supplemented by braiding of the other type of quasiparticles. 


\RBc{It is worth emphasizing that our main findings above do not contradict with the well-known fact that braiding of Majorana fermions or braiding of parafermions alone does not lead to universality. Indeed, the key observation here is that braiding of parafermions (respectively, Majorana fermions), when expressed in terms of Majorana fermions (respectively, parafermions) through our mapping, cannot be written in terms of braidings of Majorana fermions (respectively, parafermions) only. Given that Majorana fermionic braidings generate $n$-qubit Clifford gates, whereas parafermionic braidings generate $n$-qudit Clifford gates, Majorana fermionic braidings alone or parafermionic braidings alone cannot establish universality, as expected. It is the combination of Majorana fermionic braidings and parafermionic braidings that allows for universality, regardless of which representation (parafermionic or Majorana fermionic) is being used. This is made possible by the fact that one set of braidings only generates Clifford gates, whilst the other set generates some non-Clifford gates. Therefore, both perspectives indeed lead to the same conclusion, i.e., quantum universality, provided \emph{both} Majorana fermionic braidings and parafermionic braidings could be achieved in the chosen perspective. On the other hand, the presence of only one family of braidings, i.e., Majorana fermionic braidings alone or parafermionic braidings alone, is not sufficient for universality in either representation, as it should be.}

The mapping developed in this work could manifest itself in a physical system comprising one type of quasiparticle through the presence of suitable interaction terms that effectively bind the quasiparticles to form the other type of quasiparticles. \RBc{A periodically driven system is expected to be a promising platform for this purpose.} For example, the system developed in Ref.~\cite{Bomantara2021} describes two chains of Majorana fermions which are subject to appropriately devised periodic driving and Hubbard-like interaction terms, so that it supports a pair of $\mathbb{Z}_4$ parafermion edge modes. In the absence of the interaction terms, the system would simply reduce to two copies of Kitaev chains \cite{Kitaev2001} and instead supports two pairs of Majorana edge modes. In this case, depending on the interaction strength, one could store quantum information either in a $4$D-qudit spanned by two $\mathbb{Z}_4$ parafermion edge modes or in a two-qubit state spanned by \RBc{four Majorana} edge modes. Consequently, both braiding of Majorana fermions and parafermion edge modes could, in principle, be executed on demand by tuning the interaction strength and performing the appropriate braiding protocol.

In view of the above, explicitly carrying out the various braiding protocols in the physical system of Ref.~\cite{Bomantara2021} represents a promising direction for future work. Such braiding protocols could be based on a series of measurements developed in the present work or through an adiabatic means. In the former, each measurement could be carried out by connecting each end of the system to a terminal and measuring the parity-dependent conductance when appropriate voltage bias is applied \cite{Bomantara2020}. In the latter, a slowly varying Hamiltonian that enacts braiding could be made by tuning appropriate system parameters such as hopping, onsite potential, interaction strength, and/or superconducting pairing locally. 

\RBc{As precise control over interaction strength might be challenging to achieve in existing experiments, another interesting aspect for future studies is to devise a periodically driven system that supports $Z_4$ parafermion and Majorana edge modes simultaneously. In particular, as the coexistence of Majorana edge modes at different quasienergy excitations has been established and utilized for quantum computing applications in recent studies \cite{Bomantara2018PRB,Bomantara2020,Bomantara2020s}, it is reasonable to expect that such a Majorana-parafermionic hybrid system exists. In such a system, braiding of Majorana and parafermion modes could be potentially carried out under a fixed interaction strength.} \RBB{In this case, the Hilbert space spanned by the Majorana modes and that spanned by the parafermion modes could in principle be coupled in a controllable manner by adapting the adiabatic protocol developed in Ref.~\cite{Bomantara2018PRB}. In particular, this could be achieved by varying the adiabatic parameters every appropriate integer multiple of the driving period, so that the Hilbert spaces spanned by different quasienergy excitations are joined temporarily during the adiabatic process; the Hilbert spaces remain independent outside the application of the protocol, thereby allowing subsequent braiding operations to be carried out without unforeseen complications, whilst quantum information could be transferred from one Hilbert space (e.g., spanned by Majorana fermions) to another (e.g., spanned by parafermions) as needed.}

\RBB{That said, it is worth noting that the aforementioned adiabatic protocol is geometric in nature and not strictly topological, i.e., it may be sensitive to certain types of local noise or parameter fluctuations. Nevertheless, it remains more robust than typical dynamical approaches because it does not require precise control over protocol duration. We also acknowledge that the protocol of Ref.~\cite{Bomantara2018PRB} is applied to the case where the two Hilbert spaces are spanned by Majorana fermions. Therefore, while it should be feasible to extend such a protocol to the case where one Hilbert space is spanned by Majorana fermions and the other by parafermions, the detail of its construction warrants a separate study that is beyond the scope of this present work.} 

\RBB{To conclude the above discussion, our findings could in principle be implemented in some periodically driven Majorana-parafermionic hybrid system, the topologically protected braidings of which could be appropriately performed in the Majorana sector or parafermionic sector of the total Hilbert space in a way that does not require applying nonlocal transformations physically. However, the caveat is that to truly establish universality, an additional adiabatic protocol that is geometric (not strictly topological) in nature must be supplemented. Nevertheless, this drawback in our proposed idea is by no means a general drawback to implementing our findings. We remain optimistic that future developments in device design or theory may reveal a fully topological approach for coupling the Majorana fermionic and parafermionic subspaces. In this case, the drawback of our implementation proposal could actually serve as a promising motivation for the further development and studies of a Majorana-parafermionic hybrid system and the means to harness its universality as established in this work. We thus hope that our findings and our implementation proposal may inspire others to explore alternative routes toward this goal.}



 \begin{acknowledgements}
 This work was supported by the Deanship of Research
Oversight and Coordination (DROC) at King Fahd University of Petroleum \& Minerals (KFUPM) through project No.~EC221010, as well as by the Deanship of Research Oversight and Coordination (DROC) and the Interdisciplinary Research Center(IRC) for Intelligent Secure Systems (ISS) at King Fahd University of Petroleum \& Minerals (KFUPM) through internal research grant No.~INSS2507.
 \end{acknowledgements}

\appendix

\section{Majorana Parity has $1,-1$ Eigenvalues}
\label{app:1}
Let $P$ be a unitary $2\times 2$ matrix that squares non-trivially to identity (the Majorana parity $P_{jk}^{(MF)}$ is an example of such an operator). Since $U$ is unitary, it has a complete set of eigenstates (two eigenstates for a $2\times 2$ matrix). We can explicitly construct the two orthogonal eigenstates of $P$ as follows. Let $\ket{x}$ be a generic state that is not an eigenstate of $P$. Then,  $\ket{1}$=$\ket{x}+P\ket{x}$ is an eigenstate of $P$ with an eigenvalue of $1$ and $\ket{-1}$=$\ket{x}-P\ket{x}$ is an eigenstate of P with eigenvalue of $-1$. Note that $\ket{\pm1}\neq 0$ since $\ket{x}$ is not an eigenstate of $P$. Moreover, they are orthogonal to each other. \qed

\section{$\mathbb{Z}_4$ Parafermion Parity has $1,i,-1,-i$ Eigenvalues}
\label{app:2}
Let $P$ be a unitary $4\times 4$ matrix that has the property of $P^4=\IM$ non-trivially (the parafermionic parity $P_{jk}^{(P)}$ is an example of such an operator). Since $U$ is unitary, it has a complete set of eigenstates (four eigenstates for a $4\times 4$ matrix). We can explicitly construct the two orthogonal eigenstates of $P$ as follows. Let $\ket{x}$ be a generic state that is not an eigenstate of $P+P^2+P^3$ or $P-P^2+P^3$ or $iP-P^2-iP^3$ or $-iP-P^2+iP^3$  . Then,  $\ket{\lambda}=\ket{x}+\lambda^3 P\ket{x}+\lambda^2 P^2\ket{x}+\lambda P^3\ket{x}$ is an eigenstate of $P$ with eigenvalue  $\lambda\in \{\pm 1,\pm i\}$. Note that $\ket{\pm1,\pm i}\neq 0$ since $\ket{x}$ is not an eigenstate of $P+P^2+P^3$ or $P-P^2+P^3$ or $iP-P^2-iP^3$ or $-iP-P^2+iP^3$. Moreover, they are mutually orthogonal to one another. \qed 

\section{Braiding operators for parafermions do not simply exchange parafermion operators.} 
\label{app:A}

While we explicitly consider the case of $\mathbb{Z}_4$ parafermions here, the argument presented below could be straightforwardly extended to the more general case of $\mathbb{Z}_n$ parafermions. Assume for the sake of contradiction that there exists a unitary operator such that : 
\begin{equation} U^\dagger \psi_1 U=c\psi_2 \quad  \& \quad  U^\dagger \psi_2 U=c' \psi_1   
\end{equation}
 \\ where $c$ and $c’$ are constants.
Given that $U$ is unitary, the two equations imply that :
\begin{equation}
    U=c\psi_1^3U\psi_2 \label{A2}
\end{equation}
    and
    \begin{equation}
    U=c'\psi_2^3U\psi_1 \label{A3}
\end{equation}
Consider a basis of $4$ dimensional operators consisting of  all linearly independent multiplicative combinations of $\psi_1$ and $\psi_2$.
Let us further assume that $U$ contains an identity term. In this case, Eq.~(\ref{A2}) and the fact that $\psi_1^4=\psi_2^4=I$ imply that $U$ contains $A=s(I+c\psi_1^3\psi_2+c^2\psi_1^2\psi_2^2+c^3\psi_1\psi_2^3)$ and $c^4=1$. Assuming $U=A$ and Substituting in Eq.~(\ref{A3}) gives
\[
\begin{split}
     s(I+c\psi_1^3\psi_2+c^2\psi_1^2\psi_2^2+c^3\psi_1\psi_2^3)=c' s(i\psi_1\psi_2^3-icI\\+ic^2\psi_1^3\psi_2-ic^3\psi_1^2\psi_2^2) .
\end{split}
\] 
This implies that :
\begin{eqnarray}
   ic'=c^3 & \& & -icc'=1 
\end{eqnarray}
since $c^4=1$ , this implies 
\begin{eqnarray}
   ic'c=1 & \& & -icc'=1 
\end{eqnarray}
which is a contradiction. To complete the proof, one should consider the three remaining cases of $U$ containing $\psi_1,$ $\psi_1^2$, and $\psi_1^3$. Repeating the same steps as above, all of them leads to a similar inconsistent set of equations.

\section{Braiding of Majorana fermions under a conserved total parity}
\label{app:Majbraid}
In a system of \RBc{six Majorana} fermions where the total parity is conserved to $1$, i.e. the states are restricted to the $+1$ eigenstates of the total parity, the action of the total parity operator is identical to the identity. Hence, 
\begin{equation}
P_{12}^{(MF)}P_{34}^{(MF)}P_{56}^{(MF)}=\IM . \label{cparmf}
\end{equation}
In terms of the parity operators, the braiding operators read, 
\begin{equation}
  U_{ij}^{(MF)}=\frac{\IM+\gamma_i\gamma_j}{\sqrt{2}}=\frac{\IM-\mathrm{i}P_{ij}^{(MF)}}{\sqrt{2}}  \label{braidmfpar}
\end{equation}
By noting that
\begin{eqnarray*}
\sigma_z^{(1)}=P_{12}^{(MF)} &,& \sigma_x^{(1)}=P_{13}^{(MF)} , \\
\sigma_z^{(2)}=P_{45}^{(MF)} &,& \sigma_x^{(2)}=P_{46}^{(MF)} ,
\end{eqnarray*}
and further invoking Eq.~(\ref{cparmf}), i.e.,
\begin{equation*}
P_{34}^{(MF)}=P_{12}^{(MF)}P_{56}^{(MF)} \;\text{($\because \left(P_{ij}^{(MF)}\right)^2=\IM$)}
\end{equation*}
we obtain
\begin{eqnarray*}
P_{56}=i\gamma_5\gamma_6=-i\gamma_4\gamma_5\gamma_4\gamma_6=iP_{45}P_{46}=i\sigma_z^{(2)}\sigma_x^{(2)}=-\sigma_y^{(2)}  \\
P_{23}=i\gamma_2\gamma_3=-i\gamma_1\gamma_2\gamma_1\gamma_3=iP_{12}P_{13}=i\sigma_z^{(1)}\sigma_x^{(1)}=-\sigma_y^{(1)}  \\ 
\therefore P_{34}=-\zo\ys . 
\end{eqnarray*}
Also, if $k>j$, 
\begin{equation}
  \gamma_j\gamma_k=\prod_{l=j}^{l=k-1}\gamma_l\gamma_{(l+1)}=\prod_{l=j}^{l=k-1}-iP_{l(l+1)}=i^{3(k-j)}\prod_{l=j}^{l=k-1}P_{l(l+1)}  
\end{equation}
The above equations lead to
\begin{eqnarray*}
-iP_{24}^{(MF)} &=& i^{6}P_{23}P_{34}=-\yo\zo\ys=-i\xo\ys \\
-iP_{14}^{(MF)}&=& i^{6}P_{12}P_{24}=-(\zo) (\xo\ys) =-i\yo\ys \\
-iP_{15}^{(MF)}&=&i^{6}P_{14}P_{45}=-(\yo\ys)\zs=-i\yo\xs \\
-iP_{16}^{(MF)}&=&i^{6}P_{15}P_{56}=-(\yo\xs)(-\ys)=i\yo\zs \\ 
-iP_{25}^{(MF)}&=&i^{6}P_{24}P_{45}=-(\xo\ys)(\zs)=-i\xo\xs \\
-iP_{26}^{(MF)}&=&i^{6}P_{25}P_{56}=-(\xo\xs)(-\ys)=i\zs\xo \\
-iP_{35}^{(MF)}&=&i^{6}P_{34}P_{45}=-(-\zo\ys)(\zs)=i\zo\xs \\
-iP_{36}^{(MF)}&=&i^{6}P_{35}P_{56}=-(-\zo\xs)(-\ys)=-i\zo\zs 
 \end{eqnarray*}
Upon plugging in the above to Eq.~(\ref{braidmfpar}), the quantum gate operations associated with the braiding operators presented in the main text are obtained. Moreover, the quantum gate operations associated with braiding the remaining pairs of Majorana fermions could be explicitly obtained as follows,
\begin{eqnarray*}
 U_{56}^{(MF)}&=&\frac{\IM-iP_{56}^{(MF)}}{\sqrt{2}}=\frac{\IM+i\sigma_y^{(2)}}{\sqrt{2}}=\sigma_z^{(2)}\cdot H^{(2)} \\ 
U_{15}^{(MF)} &=&  CNOT_{1,2} \cdot H^{(1)} \cdot CNOT_{1,2} \cdot \sigma_z^{(1)}  \\
U_{16}^{(MF)} &=&  \sigma_z^{(1)} \cdot H^{(1)} \cdot CNOT_{2,1} \cdot H^{(1)} \cdot CNOT_{2,1}  \cdot H^{(1)}  \\
U_{25}^{(MF)} &=&  (S^{(2)})^3 \cdot CNOT_{2,1} \cdot H^{(2)} \cdot CNOT_{2,1} \cdot (S^{(2)})^3  \\
U_{26}^{(MF)} &=&   S^{(1)} \cdot H^{(1)} \cdot CNOT_{2,1} \cdot H^{(1)}  \cdot CNOT_{2,1} \\ &\cdot& H^{(1)} \cdot S^{(1)}  \\
U_{35}^{(MF)} &=& - (S^{(2)})^3 \cdot H^{(2)} \cdot \sigma_z^{(1)} \cdot \sigma_z^{(2)} \cdot H^{(1)} \cdot CNOT_{2,1}  \\ &\cdot&   H^{(1)} \cdot CNOT_{1,2} \cdot \sigma_x^{(2)} \cdot (S^{(2)})^3 \\
U_{36}^{(MF)} &=& \frac{(1 - i)}{\sqrt{2}} S^{(1)} \cdot S^{(2)} \cdot H^{(1)} \cdot CNOT_{2,1} \cdot H^{(1)}  \\
U_{45}^{(MF)} &=&  \frac{(1 - i)}{\sqrt{2}} \cdot S^{(2)}\\
U_{46}^{(MF)} &=& (S^{(2)})^3\cdot H^{(2)} \cdot (S^{(2)})^3
\end{eqnarray*}
If the system is instead projected on the $-1$ subspace of the total parity, Eq.~(\ref{cparmf}) becomes 
\begin{equation*}
  P_{12}^{(MF)}P_{34}^{(MF)}P_{56}^{(MF)}=-\IM ,
\end{equation*}
and the same procedure above can be repeated to obtain the quantum gate operations associated with all the braiding operators. In this case, a relative minus sign may arise in Eq.~(\ref{braidmfpar}) for some pairs of Majorana fermions, thus resulting in a slightly different set of quantum gate operations.

\section{Braiding of parafermions under a conserved total parity}
\label{app:Pfbraid}
In a system of four $\ZM_4$ parafermions where the total parity is conserved to $1$, i.e. the states are restricted to the $+1$ eigenstates of the total parity, the action of the total parity operator is identical to the identity. Hence, 
\begin{equation}
    P_{12}^{(P)}P_{34}^{(P)}=\IM . \label{cparpf}
\end{equation}
Recall that the braiding operator involving $\psi_i$ and $\psi_j$ is given by Eq.~(\ref{braidpf}) in the main text, which can be written in terms of parefermion parities as
\begin{equation}
  U_{ij}^{(P)}=\frac{\IM+i^{-3/2}P_{ij}^{(P)}-(P_{ij}^{(P)})^2+i^{-3/2}(P_{ij}^{(P)})^3}{2}  \label{braidpfpar}
\end{equation}
By noting that
\begin{eqnarray*}
       \Tau_z=P_{12}^{(P)} &,& \Tau_x=P_{13}^{(P)} ,
\end{eqnarray*}
    and further invoking Eq.~(\ref{cparpf}), i.e.,
    \begin{equation*}
        P_{34}^{(P)}=(P_{12}^{(P)})^3=\Tau_z^3 \;\text{($\because (P_{ij}^{(P)})^4=\IM$)}
    \end{equation*}
    we obtain
\begin{eqnarray*}
P_{23}^{(P)}&=&i^{3/2}\psi_3\psi_2^3=i^{3/2}\psi_3\psi_1^3\psi_1\psi_2^3=i^{3/2} P_{13}^{(P)}(P_{12}^{(P)})^3\\ &=& i^{3/2}\Tau_x\cdot \Tau_z^3 \\ P_{24}^{(P)}&=&i^{3/2}\psi_4\psi_2^3=i^{3/2}\psi_4\psi_3^3\psi_3\psi_2^3=i^{-3/2}P_{34}^{(P)}P_{23}^{(P)}\\
     &=&\Tau_z^3 \cdot \Tau_x\cdot \Tau_z^3=-i\Tau_z^2 \cdot \Tau_x \\ P_{14}^{(P)}&=&i^{3/2}\psi_4\psi_1^3=i^{3/2}\psi_4\psi_2^3\psi_2\psi_1^3=i^{-3/2}P_{24}^{(P)}P_{12}^{(P)} \\
     &=&i^{3/2}\Tau_z^2 \cdot \Tau_x \cdot \Tau_z=i^{-1/2}\Tau_x \cdot \Tau_z^3
     \end{eqnarray*}
     The quantum gate operations arising from each braiding operator can therefore be obtained by plugging in the above equations to Eq.~(\ref{braidpfpar}).
     
For the other cases of projecting the system on the $s \in \{1,i.-1,-i\}$ subspace of the total parity, Eq.~(\ref{cparpf}) becomes 
\begin{equation*}
  P_{12}^{(P)}P_{34}^{(P)}=s\IM ,
\end{equation*}
and the same procedure above can again be repeated to obtain the slightly different set of quantum gate operations associated with all the braiding operators. 


\section{A series of measurements for braiding parafermions}
\label{app:pmeasure}

We start by writing the explicit expression for each measurement operator presented in the main text, i.e., 
\begin{widetext}
    \begin{equation}
        M = \frac{1}{4^4} (\IM + s_1^3 P_{34} + s_1^2 P_{34}^2 + s_1 P_{34}^3) (\IM + s_2^3 P_{23} + s_2^2 P_{23}^2 + s_2 P_{23}^3) (\IM + s_3^3 P_{13} + s_3^2 P_{13}^2 + s_3 P_{13}^3) (\IM + s_4^3 P_{34} + s_4^2 P_{34}^2 + s_4 P_{34}^3) ,
    \end{equation}
\end{widetext}
To simplify the above expression, we first expand the terms within the two middle brackets by carrying out the full multiplications, thus resulting in $16$ terms within the middle bracket. By exploiting the fact that 
\begin{eqnarray}
  P_{13}^{(P)} P_{34}^{(P)} = iP_{34}^{(P)} P_{13}^{(P)} &,& P_{23}^{(P)} P_{34}^{(P)} = iP_{34}^{(P)} P_{23}^{(P)} , \nonumber \\
\end{eqnarray}
we move the terms within the last bracket next to the first bracket, which in doing so splits the terms within the middle bracket into four groups as follows, 
\begin{widetext}
    \begin{eqnarray}
        M &=& \frac{1}{4^4}[ (\IM + s_1^3 P_{34} + s_1^2 P_{34}^2 + s_1 P_{34}^3) (\IM + s_4^3 P_{34} + s_4^2 P_{34}^2 + s_4 P_{34}^3)(\IM + s_2^3 s_3 P_{23} P_{13}^3 + s_2^2 s_3^2 P_{23}^2 P_{13}^2 + s_2 s_3^3 P_{23}^3 P_{13}) \nonumber \\
        &+& (\IM + s_1^3 P_{34} + s_1^2 P_{34}^2 + s_1 P_{34}^3) (\IM + is_4^3 P_{34} + i^2 s_4^2 P_{34}^2 + i^3 s_4 P_{34}^3) (s_2^3 P_{23} + s_3^3 P_{13} + s_2^2 s_3 P_{23}^2 P_{13}^3 + s_2 s_3^2 P_{23}^3 P_{13}^2) \nonumber \\
        &+&  (\IM + s_1^3 P_{34} + s_1^2 P_{34}^2 + s_1 P_{34}^3)
(\IM + i^2 s_4^3 P_{34} + s_4^2 P_{34}^2 + i^2 s_4 P_{34}^3) (s_2^2 P_{23}^2 + s_3^2 P_{13}^2 + s_2 s_3 P_{23}^3 P_{13}^3 + s_2^3 s_3^3 P_{23} P_{13}) \nonumber \\
&+& (\IM + s_1^3 P_{34} + s_1^2 P_{34}^2 + s_1 P_{34}^3)
(\IM + i^3 s_4^3 P_{34} + i^2 s_4^2 P_{34}^2 + is_4 P_{34}^3) (s_2 P_{23}^3 + s_3 P_{13}^3 + s_2^2 s_3^3 P_{23}^2 P_{13} + s_2^3 s_3^2 P_{23} P_{13}^2) ] \nonumber \\
    \end{eqnarray}
\end{widetext}
Finally, we use $p_{ij}^{(P),s} p_{ij}^{(P),s'}=\delta_{s,s'}p_{ij}^{(P),s'}$ to further simplify the first two brackets and finally obtain,
\begin{widetext}
    \begin{eqnarray}
        M &=& \frac{1}{4^3}[(\IM + s_1^3 P_{34} + s_1^2 P_{34}^2 + s_1 P_{34}^3)
(\IM + s_2^3 s_3 P_{23} P_{13}^3 + s_2^2 s_3^2 P_{23}^2 P_{13}^2 + s_2 s_3^3 P_{23}^3 P_{13}) \delta_{s_1,s_4} \nonumber \\
&+& (\IM + s_1^3 P_{34} + s_1^2 P_{34}^2 + s_1 P_{34}^3)
(s_2^3 P_{23} + s_3^3 P_{13} + s_2^2 s_3 P_{23}^2 P_{13}^3 + s_2 s_3^2 P_{23}^3 P_{13}^2) \delta_{s_1,-is_4} \nonumber \\
&+& (\IM + s_1^3 P_{34} + s_1^2 P_{34}^2 + s_1 P_{34}^3)
(s_2^2 P_{23}^2 + s_3^2 P_{13}^2 + s_2 s_3 P_{23}^3 P_{13}^3 + s_2^3 s_3^3 P_{23} P_{13}) \delta_{s_1,-s_4} \nonumber \\
&+& (\IM + s_1^3 P_{34} + s_1^2 P_{34}^2 + s_1 P_{34}^3)
(s_2 P_{23}^3 + s_3 P_{13}^3 + s_2^2 s_3^3 P_{23}^2 P_{13} + s_2^3 s_3^2 P_{23} P_{13}^2) \delta_{s_1,is_4} ]
    \end{eqnarray}
\end{widetext}
At specific values of $s_1$ and $s_4$, only one out of the four terms in the above equation will survive due to the presence of the Kronecker delta factors. By further fixing $s_2$ and $s_3$, the terms in the second bracket will reduce to one of the unitary operators presented in the main text that act on the subspace spanned by $\psi_1$ and $\psi_2$. One of these operators is the exact braiding operator involving $\psi_1$ and $\psi_2$, whereas the rest can be written as the exact braiding operator followed by some four-dimensional Clifford gates.

\end{document}